\documentclass[a4paper]{esannV2}
\usepackage[utf8]{inputenc}
\usepackage[T1]{fontenc}
\usepackage[american]{babel}
\usepackage{amsmath}
\usepackage{graphicx}
\usepackage{tikz}
\usepackage[hidelinks]{hyperref}
\newsavebox{\architecturecheck}
\begin{document}
\title{Block Search Conditioned on Features for Sequential Quantum Data Encoding}

\author{Cecília S. C. Botelho$^{1,2}$, Erico S. Teixeira$^2$, Anderson F. P. dos Santos$^2$, Guilherme D. Lima$^1$, \\Helida Santos$^3$, Adenauer Yamin$^1$, Renata Reiser$^1$ and Inês de Castro Dutra$^4$
\vspace{.3cm}\\
1- Universidade Federal de Pelotas - CDTec \\
Rua Gomes Carneiro, 1, Pelotas--RS, Brazil
\vspace{.1cm}\\
2- Venturus - Centro de Excelência em Computação Quântica \\
Estrada Giuseppina Vianelli di Napolli, 1185 - Brazil
\vspace{.1cm}\\
3- Universidade Federal do Rio Grande - C3 \\
Av. Itália km 08 Campus Carreiros, 96201-900, Rio Grande--RS, Brazil
\vspace{.1cm}\\
4- Departamento de Ciência de Computadores \\
RISE Health Research Network, Porto, Portugal
}

\maketitle

\begin{abstract}
Sequential quantum data encoding is addressed through the joint selection of input features and encoding blocks within a fixed quantum register. At each step, a new feature, a block of operations, its placement, and its rotation weights are selected together from training data, extending the incremental construction of Q-FLAIR. With direct-loss selection (RAW), nine to twelve features were retained in four qubits across five new Bars and Stripes datasets. Under the same final calibration, mean changes relative to the calibrated Q-FLAIR reference were +5.53 percentage points in balanced accuracy, $-15.83\%$ in binary cross-entropy (BCE), and $+7.79\%$ in area under the ROC curve (AUC). The prespecified success criterion was met in four of five datasets and for the mean paired differences. These results were obtained by comparing complete procedures under a common cap of 32 operations.
\end{abstract}

\section{Introduction}

Classical-to-quantum encoding influences model expressivity and resource requirements~\cite{schuld2021}. Sequential encoding allows several features to be processed within a fixed register, offering a way to increase input capacity without increasing the number of qubits~\cite{perez2020}. This makes the choice of encoding operations and their placement relevant under a limited circuit-depth budget. Joint optimization of circuit structure and parameters has been studied~\cite{ostaszewski2021}, including architecture search with hierarchical circuit representations~\cite{lourens2023}. EAQAS combines embedding and gate structure through Bayesian optimization and cross-attention~\cite{dai2026}.

We extend Q-FLAIR's incremental joint selection of a gate, feature, and weight~\cite{qflair} to a two-operation block associated with a new feature. Block type, placement, and both rotation weights are selected jointly before insertion. The resulting decision couples feature selection to the current circuit: a feature is evaluated together with the two operations that encode it, rather than selecting the operations independently. 

\section{Block construction by input feature}
\subsection{Encoding blocks}
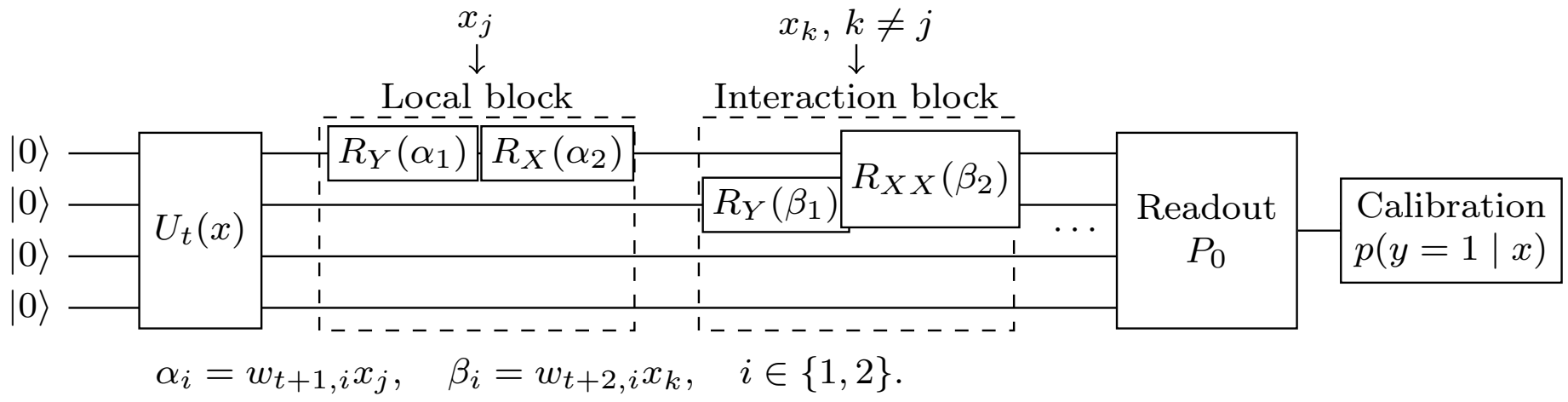
\begin{figure}[ht]
\centering
\begin{lrbox}{\architecturecheck}
\begin{tikzpicture}[x=1cm,y=.8cm,font=\footnotesize,
 gate/.style={draw,fill=white,minimum width=1cm,minimum height=.4cm,inner sep=2pt},
 every path/.style={line width=.45pt}]
\foreach \y in {0,.5,1,1.5}{
  \draw (0,\y)--(9.1,\y);
  \node[anchor=east] at (0,\y) {$|0\rangle$};
}
\draw[fill=white] (.55,-.2) rectangle (1.5,1.7);
\node at (1.025,.75) {$U_t(x)$};
\draw[dashed] (1.95,-.22) rectangle (4.4,1.85);
\draw[dashed] (4.9,-.22) rectangle (7.35,1.85);
\node[gate] at (2.6,1.5) {$R_Y(\alpha_1)$};
\node[gate] at (3.8,1.5) {$R_X(\alpha_2)$};
\node[gate] at (5.5,1) {$R_Y(\beta_1)$};
\node[gate,minimum height=.75cm] at (6.7,1.25) {$R_{XX}(\beta_2)$};
\node[fill=white,inner sep=2pt] at (7.85,.75) {$\cdots$};
\node[draw,fill=white,align=center,minimum width=1.4cm,
 minimum height=1.52cm,inner sep=4pt] (readout) at (8.85,.75) {Readout\\$P_0$};
\draw[->] (readout.east)--(9.95,.75);
\node[draw,fill=white,align=center,minimum width=1.55cm,minimum height=.8cm,inner sep=3pt] at (10.75,.75)
 {Calibration\\$p(y=1\mid x)$};
\node at (3.175,2.06) {Local block};
\node at (6.125,2.06) {Interaction block};
\node at (3.175,2.75) {$x_j$};
\node at (6.125,2.75) {$x_k$, $k\ne j$};
\draw[->] (3.175,2.56)--(3.175,2.27);
\draw[->] (6.125,2.56)--(6.125,2.27);
\node[anchor=west] at (.55,-.65)
 {$\alpha_i=w_{t+1,i}x_j,\quad\beta_i=w_{t+2,i}x_k,\quad i\in\{1,2\}$.};
\end{tikzpicture}
\end{lrbox}
\ifdim\wd\architecturecheck>\linewidth
  \resizebox{\linewidth}{!}{\usebox{\architecturecheck}}%
\else
  \usebox{\architecturecheck}%
\fi
\caption{A local block followed by an interaction block. The sequence is selected during training and fixed for inference.}
\label{fig:architecture}
\end{figure}

For $x\in[0,\pi]^{16}$, the circuit after $t$ blocks is extended as
\begin{equation}
 U_{t+1}(x)=B_{\ell}(w_{t+1,1}x_j,w_{t+1,2}x_j)U_t(x),
 \qquad j\notin S_t,
\end{equation}
with $U_0=I$, $S_0=\emptyset$, and $S_{t+1}=S_t\cup\{j\}$. Here, $x$ is the normalized input feature vector, $U_t(x)$ is the circuit after $t$ blocks, $\ell$ specifies block type and placement, and $S_t$ contains the indices of previously selected features. The weights $w_{t+1,1}$ and $w_{t+1,2}$ multiply the same new feature $x_j$. Below, $w_1$ and $w_2$ denote these weights with the stage index $t+1$ omitted. The sequence is fixed after training (Fig.~\ref{fig:architecture}).

The library contains local and nearest-neighbor interaction blocks, each with two operations and two weights. In local blocks, $R_Y(w_1x_j)$ is followed by $R_X(w_2x_j)$ on one qubit. In interaction blocks, $R_Y(w_1x_j)$ is applied to one endpoint before $R_{XX}(w_2x_j)$ is applied to the adjacent pair. Four local placements and six endpoint orientations are considered on the chain of qubits 0, 1, 2, and 3. Rotations are defined by $R_P(\theta)=e^{-i\theta P/2}$, with Pauli operators $P=X,Y,X\otimes X$. In matrix products, the rightmost operation is applied first; thus $B_{\mathrm{loc}}=R_X(w_2x_j)R_Y(w_1x_j)$.

\subsection{Candidate objectives and search}
The joint probability of measuring all four qubits in zero is $P_0(x)=|\langle0000|U(x)|0000\rangle|^2$.
RAW scores candidates by training binary cross-entropy:
\begin{equation}
 \mathcal{J}_{\mathrm{RAW}}(U)=\operatorname{BCE}_{\rm train}(y,P_0(x)).
\end{equation}
where $\operatorname{BCE}(y,p)=-n^{-1}\sum_{i=1}^{n}[y_i\log p_i+(1-y_i)\log(1-p_i)]$. For RAW scoring, probabilities are clipped to $[10^{-12},1-10^{-12}]$. CAL fits logistic calibration within candidate scoring:
\begin{equation}
 \mathcal{J}_{\mathrm{CAL}}(U)=\min_{a,b}\left\{
 \operatorname{BCE}_{\rm train}\!\left(y,\sigma(a[2P_0(x)-1]+b)\right)
 +\frac{a^2}{2n_{\rm train}}\right\},
\end{equation}
with $\sigma(z)=(1+e^{-z})^{-1}$ and $C=1$~\cite{niculescu2005}. Both parameters are real; only $a$ is penalized. Logistic loss is computed from logits for numerical stability. Both variants use final calibration.

The probability response is represented in the basis $(1,\cos\theta,\sin\theta)$ for each rotation~\cite{schuld2021}. Sampling both angles at $0$ and $\pm\pi/2$ recovers nine tensor coefficients from a $3\times3$ grid. Eight new evaluations are needed because $(0,0)$ is reused; coefficients are shared across features. Candidate weights can therefore be scored from the reconstructed response without repeating circuit simulation for every starting pair. Each unused feature is screened at every placement using four fixed starting pairs: $(0.3,0.2)$, $(-0.3,0.2)$, $(0.6,-0.4)$, and $(-0.6,-0.4)$, chosen before evaluation to cover both signs and two nonzero magnitudes per weight.

The eight best candidates are refined jointly over $[-1,1]^2$ using L-BFGS-B~\cite{byrd1995}. For each candidate, the best result from refinement, the starting pair, and the identity is retained. A single block is selected greedily with deterministic tie-breaking; weights of earlier blocks remain fixed.

\subsection{Stopping and inference}
Construction stops after 16 blocks (32 operations), when improvement is at most $10^{-12}$, or when all features are used. For both variants, $p(y=1\mid x)=\sigma(a[2P_0(x)-1]+b)$ is fitted by CAL on training data at each complete block prefix. Validation BCE selects a prefix or a constant predictor equal to the training prevalence, preferring shorter prefixes on ties. This permits a shorter circuit than the one constructed, or no circuit when the constant predictor is preferred. Architecture and calibration are then fixed for inference.

\section{Experimental protocol}
\subsection{Data and evaluation stages}
Bars and Stripes $4\times4$ images were generated with Gaussian pixel noise of standard deviation 0.5, as in the reference benchmark~\cite{bowles2024,qflair}. The released training set was split into 800/200 training/validation examples, stratified to preserve class proportions; its 200 test examples were retained. Min-max scaling to $[0,\pi]$ was fitted only on training data, with clipping elsewhere.

The same \texttt{qml-benchmarks} generator adopted by Q-FLAIR was used at commit \texttt{f06e20c}. Using the original seed 42 reproduced the released labels exactly, with pixel differences at most $4.44\times10^{-15}$. Each new dataset contained 800/200/1,000 training/validation/test examples. Exact row duplication was checked within and across partitions; confirmation datasets were also checked against each other and all previously used datasets.

CAL was primary and RAW secondary in five initial datasets. RAW was fixed as primary before evaluation on five new confirmation datasets. Generation and stratified partitioning used separate, prespecified seeds. Code, data, seeds and reproduction scripts are available at \url{https://github.com/CeciliaBotelho/ESANN-2026}.

\subsection{Reference implementation and final evaluation}
Q-FLAIR commit \texttt{4f331783} was run with Qiskit 0.44.3 and scikit-learn 1.6.0, with the register size set to four qubits in both training and prediction scripts. Its ordered pool \texttt{Rxx\_x}, \texttt{Ryy\_x}, \texttt{U2\_x}, \texttt{U3\_x}, \texttt{H\_y}, 1,001 weight points, seed 12345, and code stopping threshold $10^{-8}$ were retained. Construction was capped at 32 operations.

Q-FLAIR used a constant $\pi$ channel, ring connectivity, and feature reuse; block search used chain connectivity and new features. Both methods received the same variable columns and data partitions. QF+CAL uses the same final calibration ($C=1$); validation BCE selects a prefix or constant predictor. QF+CAL considers every operation prefix; RAW considers complete block prefixes. Balanced accuracy is the mean recall over the two classes. Construction, calibration, and ROC threshold fitting use training data only. The uncalibrated final reference is descriptive.

\subsection{Criteria and partial controls}
The prespecified criterion required a balanced accuracy gain $\geq0.01$ and BCE difference $\leq0$ relative to QF+CAL. Confirmation required all five datasets to be completed, with the criterion met in at least four and for the mean paired differences. AUC and paired outcomes were reported descriptively.

Partial controls were included in the initial CAL evaluation. In SEQ, the feature and first rotation were selected before the second. In SEQ+REF, the SEQ feature and placement sequence were retained, with both weights refitted jointly. Pairs, connectivity, feature rules, and operation cap were shared; initialization, effort, and trajectories differed. A mean BCE reduction of 0.010 and lower BCE in four datasets were required per contrast.

Data integrity and calibration were checked, and circuit responses were compared with Qiskit replays. RAW confirmation metrics were independently recomputed from saved predictions. Every reference run was truncated at the 32 operation cap. Hyperparameters and seeds were fixed before test evaluation.

\section{Results}
\subsection{Initial evaluation and candidate objective}
On the released partition, CAL obtained 77.96\% balanced accuracy, 0.49891 BCE, and 0.82276 AUC, versus 72.54\%, 0.51244, and 0.75696 for QF+CAL, meeting the initial criterion. A 30-operation prefix was selected from 32 constructed operations, retaining 15 features in five local and ten interaction blocks. RAW obtained 86.26\%, 0.36306, and 0.90981, respectively.

Across the first five new datasets, mean balanced accuracy and BCE were 80.22\% and 0.461903 for CAL, compared with 76.52\% and 0.469159 for QF+CAL. The joint criterion was met in two of five datasets, against the requirement of four for replication. Secondary RAW balanced accuracy, BCE, and AUC were 85.91\%, 0.35575, and 0.91482.

Mean BCE was 0.43632 for SEQ, 0.461903 for CAL, and 0.469156 for SEQ+REF. CAL minus SEQ+REF gave a mean BCE difference of $-0.00725$, with lower BCE in two datasets, against the required reduction of 0.010 and consistency in four. These comparisons were partial controls for CAL. RAW was then tested on new datasets against QF+CAL.

\subsection{Prospective RAW confirmation}
Table~\ref{tab:confirmation} reports mean test results for the five confirmation datasets. Relative to QF+CAL, a mean balanced accuracy difference of $+5.53$ percentage points, a BCE difference of $-0.06945$ (15.83\% reduction), and an AUC difference of $+0.06561$ (7.79\% increase) were measured. The selected RAW prefixes retained nine to twelve features in 18--24 operations.

\begin{table}[ht]
\centering
\small{\begin{tabular}{lrrr}
\hline
Method & Balanced accuracy & BCE & AUC\\
\hline
Q-FLAIR final output & 78.32\% & 0.45623 & 0.84216\\
QF+CAL & 78.28\% & 0.43861 & 0.84230\\
Block search, RAW & \textbf{83.81\%} & \textbf{0.36916} & \textbf{0.90791}\\
Constant predictor & 50.00\% & 0.69331 & 0.50000\\
\hline
\end{tabular}}
\caption{\small{Means in five confirmation datasets, with matched final calibration and validation prefix selection. Reference construction was capped at 32 operations.}}
\label{tab:confirmation}
\end{table}

The criterion was met in four of five datasets and for the mean paired differences. BCE differences (RAW minus QF+CAL), in dataset order, were $-0.0827$, $-0.0977$, $+0.0337$, $-0.0608$, and $-0.1397$.
Mean CPU times were 1.94 s for RAW (including imports, calibration, and prefix selection) and 5,530 s for reference training (excluding replay and prediction). RAW used 1,040 batched training evaluations and 104 refinements per dataset, including stopping assessment. The times reflect different simulator implementations and parallelization.

\section{Discussion}
The reported differences compare complete procedures. Paired analyses added after evaluation of the five confirmation datasets gave unadjusted two-sided $t$-test $p$-values of 0.0436, 0.0737, and 0.0750 for balanced accuracy, BCE, and AUC. Exact sign-permutation tests over 32 assignments gave $p=0.125$ for each metric. The tests assume paired-difference normality and sign exchangeability, respectively. With only five datasets, statistical evidence is limited; meeting the prespecified criterion does not imply significance for every metric.

The experiments use one Bars and Stripes generator, $4\times4$ images, noise standard deviation 0.5, four qubits, and 32 operations. Connectivity, feature reuse, objectives, and optimization differ between procedures; the reference is truncated at the cap. The sequential controls were evaluated under CAL and do not establish whether joint refinement explains the RAW gains. Likewise, the CPU measurements do not isolate an algorithmic speedup from simulator and implementation differences. Matched controls are needed to separate these effects.

\section{Conclusion}

Joint selection of blocks, placements, and weights retained nine to twelve features in four qubits. RAW improved mean balanced accuracy by 5.53 percentage points, reduced BCE by 15.83\%, and increased AUC by 7.79\% relative to calibrated Q-FLAIR under the 32-operation cap. The prespecified criterion was met in four of five new datasets and for mean paired differences. Future experiments will test additional blocks and compare connectivity, feature reuse, and optimization under matched controls and fixed budgets.

\end{document}